\documentclass{article}
\usepackage[top=1.5in, bottom=1.5in, left=1.2in, right=1.2in]{geometry}
\usepackage{authblk}
\usepackage{amssymb, amsmath,framed}
\usepackage{graphicx}
\usepackage[square]{natbib}

\usepackage{bm}
\usepackage[colorlinks=true,linkcolor=blue,urlcolor=blue,citecolor=blue,anchorcolor=blue]{hyperref}
\expandafter\ifx\csname package@font\endcsname\relax\else
 \expandafter\expandafter
 \expandafter\usepackage
 \expandafter\expandafter
 \expandafter{\csname package@font\endcsname}
\fi
\def\bq{\begin{equation}}
\def\eq{\end{equation}}
\def\bqy{\begin{eqnarray}}
\def\eqy{\end{eqnarray}}

\makeatother

\begin{document}

\title{Electric sails are potentially more effective than light sails near most stars}

\author{Manasvi Lingam\thanks{Electronic address: \texttt{mlingam@fit.edu}}}
\affil{Department of Aerospace, Physics and Space Sciences, Florida Institute of Technology, Melbourne, FL 32901, USA}
\affil{Institute for Theory and Computation, Harvard University, Cambridge, MA 02138, USA}

\author{Abraham Loeb\thanks{Electronic address: \texttt{aloeb@cfa.harvard.edu}}}
\affil{Institute for Theory and Computation, Harvard University, Cambridge, MA 02138, USA}

\date{}

\maketitle

\begin{abstract}
Electric sails are propulsion systems that generate momentum via the deflection of stellar wind particles through electric forces. Here, we investigate the relative merits of electric sails and light sails driven by stellar radiation pressure for F-, G-, K- and M-type stellar systems. We show that electric sails originating near M-dwarfs could attain terminal speeds of $\sim 500$ km/s for minimal payload masses. In contrast, light sails are typically rendered ineffective for late-type M-dwarfs because the radiation pressure is not sufficiently high to overcome the gravitational acceleration. Our analysis indicates that electric sails are better propulsion systems for interplanetary travel than light sails in proximity to most stars. We also delineate a method by which repeated encounters with stars might cumulatively boost the speeds of light sails to $\gtrsim 0.1\,c$, thereby making them more suitable for interstellar travel. This strategy can be effectuated by reaching $\sim 10^5$ stars over the span of $\sim 10$ Myr.
\end{abstract}

\section{Introduction} \label{SecIntro}
The past decade has witnessed many advances in space exploration, planetary science and astrobiology. They include a rapid growth in the number of detected exoplanets \citep{Per18} as well as improvements in our understanding of potentially habitable worlds within our Solar system \citep{DPL15}. There has also been a renewed interest in space exploration on different fronts. Aside from the upcoming robotic missions to various objects in our Solar system, there are plans underway in both the public and private sectors to dispatch humans to the Moon and Mars within the next decade.\footnote{\url{https://www.nasa.gov/feature/sending-american-astronauts-to-moon-in-2024-nasa-accepts-challenge}} Looking beyond our Solar system neighborhood, the ambitious goal of the recently initiated \emph{Breakthrough Starshot} project is to send a gram-scale spacecraft to Proxima Centauri at the speed of $0.2\,c$ \citep{Mer16,ML17,ADI18,Park18}.\footnote{\url{https://breakthroughinitiatives.org/initiative/3}}

Space exploration has been argued to confer a number of concrete and intangible benefits in the context of both interplanetary and interstellar travel \citep{Craw09,Craw10}. There has been much debate as to which propulsion systems are well-suited for space exploration, especially of the interstellar kind \citep{MM89,Gil04,Long}. Due to the constraints imposed by the rocket equation, the focus has been increasingly shifting toward propulsion systems that do not have to carry their fuel onboard the spacecraft. The best known examples in this category are light sails propelled by electromagnetic radiation, either derived from the host star (i.e., black body radiation) or using laser arrays \citep{McIn99,FSE16,MLS17}. Over the past few decades, however, promising alternatives such as magnetic sails \citep{ZA91} and electric sails \citep{Jan04} have been formulated.

In this paper, we address the characteristics of electric sails and light sails powered by stars as a function of stellar parameters (mass and rotation rate). Although ignored here, the effectiveness of magnetic sails might be comparable to electric sails in certain respects \citep{PH16}. Broadly speaking, the electric sail propulsion system relies upon electric forces (distributed over a wire mesh) to deflect stellar wind protons, and this process consequently imparts momentum to the spacecraft and accelerates it in the direction of the wind. The concept of electric sails was first elucidated by Ref. \cite{Jan04} and this field has progressed on both the experimental and theoretical fronts in the ensuing years \citep{JTP10,QM10,SRK13,JMP15,TJ17,BMQ19}.

The outline of the paper is as follows. We begin by describing how the basic features of electric sails are regulated by stellar parameters in Sec. \ref{SecESail}. We follow this up in Sec. \ref{SecEPS} by comparing the characteristics of electric sails and light sails for stars of different masses. We also describe how electric sails may enable the attainment of relativistic speeds and how they can be utilized for interstellar travel and exploration. Finally, we summarize our central conclusions in Sec. \ref{SecConc}.

\section{Deploying electric sails near other stars}\label{SecESail}
We will commence our analysis by exploring how the basic characteristics of electric sails are dependent on the properties of their host star. 

\subsection{Preliminary considerations}
We begin by introducing some essential nomenclature. The distance from the star where the stellar flux equals the solar flux incident upon the Earth is denoted by $d_\star$. In terms of the stellar luminosity ($L_\star$), it is thus given by
\begin{equation}\label{EqDist}
    d_\star = 1\,\mathrm{AU}\,\sqrt{\frac{L_\star}{L_\odot}},
\end{equation}
and we utilize the simple mass-luminosity relationship $L_\star \propto M_\star^3$ for low-mass stars with $M_\star \lesssim 2 M_\odot$ \citep{BV92,SC05}, which allows us to express $d_\star$ in terms of the stellar mass ($M_\star$). For example, if we consider Proxima Centauri, this scaling yields $L_\star \approx 1.8 \times 10^{-3}\,L_\odot$, which is very close to $L_\star = 1.55 \times 10^{-3}\,L_\odot$ inferred from observations \citep{BBB12}.

In this paper, we will posit that technological entities, irrespective of whether they are extraterrestrial technological species or \emph{Homo sapiens}, may prefer to launch or maneuver interstellar spacecrafts from locations close to $d_\star$. A similar hypothesis was advanced by Ref. \cite{LL18b} while studying the feasibility of chemical propulsion around low-mass stars. There are several reasons as to why $d_\star$ could constitute an optimal choice. First, at distances much closer than $d_\star$, the spacecrafts will be subject to higher heating as well as more susceptible to deleterious space weather events such as coronal mass ejections. Second, as we shall see later, distances far greater than $d_\star$ will yield weaker accelerations and lower terminal speeds for certain spacecrafts. Lastly, if any exoplanets exist in the habitable zone - which would be situated at distances similar to $d_\star$ \citep{Dole,KWR93,KRF13,RAF19} - they are well-suited to serve as ``natural'' bases for launching these spacecrafts.

\subsection{Stellar wind pressure and radiation pressure}
Before embarking on our analysis, it is instructive to compare the ratio of the dynamical pressure exerted by the stellar wind ($P_\mathrm{wind}$) and the stellar radiation pressure ($P_\mathrm{rad}$). For a spherically symmetric stellar wind, we have
\begin{equation}\label{DensRel}
    \rho = \frac{\dot{M}_\star}{4\pi r^2 v},
\end{equation}
where $\dot{M}_\star$ is the stellar mass-loss rate, whereas $\rho$ and $v$ are the mass density and the velocity of the solar wind, respectively, at the distance $r$. In this paper, we shall consider the deployment of electric (or solar) sails at distances comparable to $d_\star$ or larger. At such distances, the stellar wind can be assumed to have approximately constant speed \citep{GV18}. Furthermore, the value of this asymptotic speed is close to the escape speed from the star \citep{CS11}, which coincidentally does not vary greatly across stars \citep{JGL15}. Hence, we shall hereafter adopt $v \approx v_0 = 500$ km/s because it lies within a factor of $2$ with respect to the observed fast and slow solar wind velocities \citep{Mar06,SN09}, as well as the values of $v$ for the seven TRAPPIST-1 planets \citep{GTD17} and Proxima b \citep{AE16} predicted by numerical simulations \citep{GDC16,DLMC,DJL18}. 

\begin{figure}
\includegraphics[width=7.5cm]{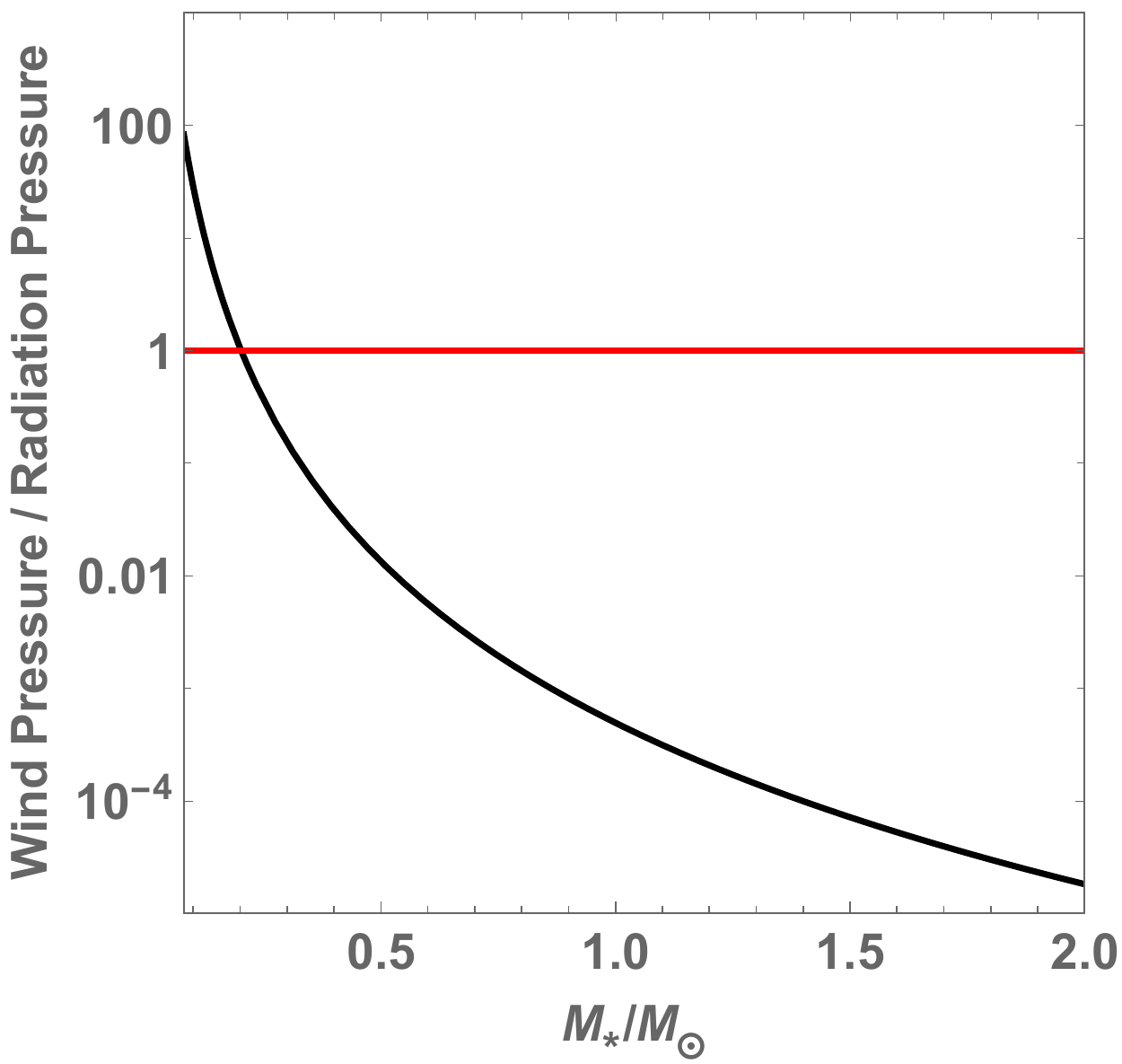} \\
\caption{The ratio of the stellar wind dynamical pressure and the stellar radiation pressure as a function of the stellar mass $M_\star$ (in units of $M_\odot$). The horizontal red line corresponds to the scenario wherein the two pressures are equal to one another.}
\label{FigPRatio}
\end{figure}

With this simplification, we note that the dynamical pressure of the stellar wind is $P_\mathrm{wind} = \rho v^2$, whereas the stellar radiation pressure is \citep{McIn99}:
\begin{equation}
 P_\mathrm{rad} = \frac{L_\star}{4\pi r^2 c}.   
\end{equation}
Thus, the ratio of these two pressures ($\delta_P$) is expressible as
\begin{equation}\label{PresRat}
    \delta_P = \frac{P_\mathrm{wind}}{P_\mathrm{rad}} \approx \frac{\dot{M}_\star v_0 c}{L_\star}. 
\end{equation}
In order to complete this expression, we must determine $\dot{M}_\star$ as a function of basic stellar parameters. However, this is not easy to undertake for the following reasons: (i) the stellar mass-loss rate is predicted to change substantially over time \citep{WMZ05,OV18,Lin19}, and (ii) the mass-loss rate scalings for low- and high-mass stars are still subject to debate \citep{SCu05,CS11}. In order to address (ii), we will restrict our attention to F-, G-, K- and M-type stars as they collectively constitute over $90\%$ of all stars in the Milky Way \citep{Krou01}. For these low-mass stars, we will make use of the mass-loss rate obtained via numerical simulations by Ref. \cite{JGB15}:
\begin{equation}\label{MLPres}
    \frac{\dot{M}_\star}{\dot{M}_\odot} \approx \left(\frac{\Omega_\star}{\Omega_\odot}\right)^{1.33} \left(\frac{M_\star}{M_\odot}\right)^{-1.76},
\end{equation}
where $\dot{M}_\odot \approx 2 \times 10^{-14}\,M_\odot\,\mathrm{yr}^{-1}$ is the current solar mass-loss rate, while $\Omega_\star$ and $\Omega_0$ denote the rotation rate of the star and the Sun, respectively. Note, however, that (\ref{MLPres}) is predicted to be accurate mainly in the range $0.4 < M_\star/M_\odot < 1.1$ \citep{JGB15}. 

For late-type M-dwarfs, which are fully convective, the validity of this model is not guaranteed. For instance, if we substitute the parameters for Proxima Centauri (whose mass is $M_\star = 0.12 M_\odot$) into (\ref{MLPres}), we find that $\dot{M}_\star \approx 5 \dot{M}_\odot$ \citep{LL18a}. This estimate is compatible with the $3\,\sigma$ upper limit of $\dot{M}_\star \approx 14 \dot{M}_\odot$ for this star based on X-ray constraints \citep{WD02}. On the other hand, numerical simulations of the mass-loss rate have yielded $\dot{M}_\star \approx \dot{M}_\odot$ \citep{GDC16}. It is important to recognize that the presence of $\Omega_\star$ in (\ref{MLPres}) implies that $\dot{M}_\star$ is implicitly dependent on the stellar age because $\Omega_\star$ evolves with time. Moreover, the issue of calculating the time-dependent mass-loss rate is complicated by the fact that active young stars may lose a non-negligible fraction of their mass via coronal mass ejections \citep{DCY13,Cran17,LL19}. 

In our subsequent analysis, to simplify the inherent complexity, we will adopt $\Omega_\star \sim \Omega_\odot$ and assume that (\ref{MLPres}) is applicable over the broader mass range of $0.075 \leq M_\star/M_\odot \leq 2$. Extending this domain to encompass convective M-dwarfs implicitly amounts to supposing that these stars are very active. The assumption concerning the rotation rate implicitly ignores the time-dependent evolution of the rotation rate \citep{Sku72,SZ08}. In particular, for late-type M-dwarfs that are predicted to have total lifetimes of $\gtrsim 1000$ Gyr, the mass-loss rates towards the end of their main-sequence stage will be much lower than their present-day values because the mass-loss rate declines sharply over time \citep{Lin19,WMZL,DLM18}. However, our results are not altered significantly since we are primarily interested in analyzing the prospects for propulsion systems in the past and near-future; a substantial fraction of M-dwarfs in the current epoch exhibit rotation rates not far removed from $\Omega_\odot$ \citep{NMI18}. We do not therefore tackle the feasibility of different propulsion systems in the cosmic distant future (corresponding to $\gtrsim 100$ Gyr) because it is impossible to predict technological evolution on such timescales.

After implementing the above scalings, we find that (\ref{PresRat}) reduces to
\begin{equation}\label{FinPRat}
    \delta_P \sim 4.9 \times 10^{-4}\,\left(\frac{M_\star}{M_\odot}\right)^{-4.76}.
\end{equation}
By inverting this expression, it is straightforward to verify that $\delta_P \gtrsim 1$ is attained when $M_\star \lesssim 0.2 M_\odot$. Thus, for late-type M-dwarfs, it is conceivable that the dynamical pressure from the stellar wind exceeds the radiation pressure. Note that (\ref{FinPRat}) does not exhibit any dependence on the distance from the star and the stellar age (or equivalently $\Omega_\star$) due to the simplifications introduced earlier. We have plotted $\delta_P$ as a function of $M_\star$ in Fig. \ref{FigPRatio}; it is evident that this ratio varies by several orders of magnitude across the stellar mass range studied in this paper. We reiterate that the estimate for $\delta_P$ is expected to be valid only when the stellar rotation rate is comparable to $\Omega_\odot$; it will decline by orders of magnitude (relative to the present-day value) for M-dwarfs approaching the end of their main-sequence lifetime. 

\subsection{Properties of electric sails}\label{SSecPropES}
Now, we turn our attention to the feasibility of deploying electric sails for propulsion. The force per unit length ($dF/dz$) generated by an electric sail has been estimated by Ref. \cite{JA07} and \cite{TJ09} to be
\begin{equation}\label{FbyL}
 \frac{d F}{d z} = \frac{K m_p n v^2 r_0}{\sqrt{\exp\left[\frac{m_p v^2}{e V_0}\ln(r_0/r_w)\right]-1}},  
\end{equation}
where $K \approx 3.09$ is a dimensionless constant determined via numerical simulations, $m_p$ is the proton mass, $n$ is the number density of the stellar wind, $V_0$ is the potential at which the wires comprising the light sail are maintained, $r_w$ is the wire radius and $r_0$ is defined as follows:
\begin{equation}\label{WirePar}
    r_0 = 2\sqrt{\frac{\epsilon_0 k_B T_e}{n e^2}},
\end{equation}
where $T_e$ represents the electron temperature of the stellar wind and $\epsilon_0$ is the permittivity of free space; note that the term inside the square root in (\ref{WirePar}) is the Debye length. As we have chosen to work with the ansatz $v \approx v_0$, we see that the factor $\kappa = m_p v^2/(e V_0)$ in the denominator of (\ref{FbyL}) is independent of the stellar properties. After substituting the appropriate values from Ref. \cite{JA07}, it is found that $\kappa \approx 0.15$. Given that the denominator in (\ref{FbyL}) is expressible as $\sqrt{(r_0/r_w)^\kappa - 1}$, we see that the dependence on $r_0$ is rather weak and this factor is close to unity. Hence, in the spirit of Ref. \cite{JA07}, we are free to drop this term without much loss of generality. 

In order to proceed with our calculations, we must determine how $n$ and $T_e$ scale with the distance from the star and the stellar mass. With regards to the former, we observe that $n \propto \rho$ and make use of (\ref{DensRel}), (\ref{MLPres}) and $v \sim v_0$ to arrive at
\begin{equation}\label{NumDen}
    n \sim 8 \times 10^6\,\mathrm{m}^{-3}\,\left(\frac{M_\star}{M_\odot}\right)^{-1.76} \left(\frac{r}{1\,\mathrm{AU}}\right)^{-2},
\end{equation}
where the normalization has been adopted from Ref. \cite{SN09}. At distances of $d_\star$ and higher, the scaling $n \propto r^{-2}$ appears to be quite robust as per \emph{Ulysses} spacecraft measurements \citep{LIM11}, and numerical simulations of the stellar wind for the TRAPPIST-1 system \citep{DJL18}. The dependence on $M_\star$, on the other hand, emerges through the evolving mass-loss rate given by (\ref{MLPres}). Hence, it is conceivable that (\ref{NumDen}) overestimates the number density by an order of magnitude or more when it comes to M-dwarfs because the accuracy of (\ref{MLPres}) remains uncertain for $M_\star < 0.4 M_\odot$. 

Next, we make use of the following ansatz for the electron temperature:
\begin{equation}\label{ElecT}
    T_e \sim 1.4 \times 10^5\,\mathrm{K}\, \left(\frac{r}{1\,\mathrm{AU}}\right)^{-0.5},
\end{equation}
where the scaling with respect to $r$ has been adopted from measurements performed by the \emph{Ulysses} mission \citep{LIM11} and the normalization has been taken from Ref. \cite{NRP98}. We have not introduced any mass dependence in the above equation since numerical models indicate that the wind temperature has a weak dependence on the stellar mass \citep{SJV14,JGB15}. Furthermore, if we apply the above ansatz to the TRAPPIST-1 planetary system and Proxima b, we obtain values for $T_e$ that are only a factor of $\lesssim 2$ different from numerical simulations; see Table 2 of Ref. \cite{DLMC} and Table S2 of Ref. \cite{DJL18}. After substituting (\ref{NumDen}) and (\ref{ElecT}) into (\ref{WirePar}), we end up with
\begin{equation}
    r_0 \sim 18.3\,\mathrm{m}\,\left(\frac{M_\star}{M_\odot}\right)^{0.88}\left(\frac{r}{1\,\mathrm{AU}}\right)^{0.75}.
\end{equation}

We have now assembled the various components necessary for calculating the force per unit length by utilizing (\ref{FbyL}). We are, however, more interested in two quantities: (i) the acceleration at distance $d_\star$, and (ii) the terminal velocity $v_{\infty,E}$ achievable by the electric sail. In order to calculate (i), we begin by estimating $dF/dz$. After neglecting the denominator in the right-hand-side of (\ref{FbyL}) for reasons elucidated earlier, we obtain
\begin{equation}\label{ForceL}
    \frac{dF}{dz} \sim 1.9 \times 10^{-7}\,\mathrm{N/m}\,\left(\frac{M_\star}{M_\odot}\right)^{-0.88}\left(\frac{r}{1\,\mathrm{AU}}\right)^{-1.25}.
\end{equation}
For a solar-mass star at $1$ AU, the above formula yields a thrust per unit length of $\sim 1.9 \times 10^{-7}$ N/m. Despite the numerous simplifications involved, this result still exhibits good agreement with the more complex and accurate model described in Ref. \cite{JTP10} that obtained $dF/dz \approx 5 \times 10^{-7}$ N/m for the same stellar mass and distance.

In order to determine the acceleration, we must next account for the mass per unit length (denoted by $dm/dz$). There are two essential components in the electric sail: (i) the wire mesh and (ii) the electron gun \citep{JTP10}. The role of the wire mesh is to deflect stellar wind protons, and thereby transfer momentum to the electric sail. However, as the wind also contains electrons, the purpose of the electron gun is to ``pump'' out these electrons and maintain the wires at a positive potential. For the wire mesh, we have
\begin{equation}\label{MeshMass}
    \frac{dm}{dz} = \pi r_w^2 \rho_w \approx 1.3 \times 10^{-6}\,\mathrm{kg/m},
\end{equation}
where $\rho_w$ is the mass density of the wire, and the last equality follows after adopting the parameters from Ref. \cite{JA07}. Next, for the electron gun, the corresponding value of $dm/dz$ was derived in Ref. \cite{JA07}:
\begin{eqnarray}\label{EGunMass}
   && \frac{dm}{dz} = \frac{2 e n r_w V_0}{\zeta} \sqrt{\frac{2 e V_0}{m_e}} \nonumber \\
   && \quad \quad \sim 2 \times 10^{-6}\,\mathrm{kg/m} \left(\frac{\zeta}{10\,\mathrm{W/kg}}\right)^{-1} \left(\frac{M_\star}{M_\odot}\right)^{-1.76} \left(\frac{r}{1\,\mathrm{AU}}\right)^{-2},
\end{eqnarray}
where $\zeta$ is the power per unit mass associated with the electron gun and the last equality follows after using (\ref{NumDen}) and the other input parameters from Ref. \cite{JA07}. 

There are two points that deserve to be highlighted at this juncture. First, by inspecting (\ref{EGunMass}), we see that $dm/dz \propto r^{-2}$. Hence, at sufficiently large distances, the required value of $dm/dz$ for the electron gun will fall below (\ref{MeshMass}). Second, it was estimated in Ref. \cite{Jan04} that an electron gun at $1$ AU from the Sun could function at a power per unit mass of $\sim 0.3$ W/kg, which is small in comparison to $\zeta \sim 10^3$ W/kg and $\zeta \lesssim 10^5$ W/kg attainable by sports cars and rocket engines, respectively. By utilizing an apposite high-performance engine (e.g., gas turbines) to power the electron gun, it might consequently be feasible to achieve sufficiently high values of $\zeta$ even at $r \sim d_\star$ to permit the neglect of this term. As our eventual goal is to compare the performance of light sails and electric sails, it makes sense to compare the same setup (viewed \emph{sensu lato}) for both propulsion systems. We will, therefore, tackle the scenario wherein the payload mass is comparable or negligible with respect to the mass of the electric or light sail.

In this particular limit, the acceleration experienced by an electric sail ($a_E$) is found by dividing (\ref{ForceL}) with (\ref{MeshMass}), which subsequently yields
\begin{equation}\label{AccESail}
    a_E \sim 0.15\,\mathrm{m/s^2}\,\left(\frac{M_\star}{M_\odot}\right)^{-0.88}\left(\frac{r}{1\,\mathrm{AU}}\right)^{-1.25}.
\end{equation}
The acceleration at $1$ AU for $M_\star = M_\odot$ is higher than the corresponding estimate in Ref. \cite{JA07} by roughly an order of magnitude because we neglected the denominator in the right-hand-side of (\ref{FbyL}) and the payload mass resulting from the electron gun, and we specified slightly higher values for the solar wind parameters at $1$ AU. For $r = d_\star$, the location at which the electric sail is launched, the corresponding acceleration (denoted by $a_{0,E}$) becomes
\begin{equation}\label{aintE}
    a_{0,E} \sim 0.15\,\mathrm{m/s^2}\,\left(\frac{M_\star}{M_\odot}\right)^{-2.755}
\end{equation}
From this formula, it is apparent that the initial acceleration experienced by the electric sail exhibits a strong dependence on $M_\star$. 

Lastly, we can estimate the terminal velocity achieved by the electric sail by following the procedure described in Ref. \cite{JA07}. The equation of motion for any object subjected to a spherically symmetric potential in the presence of gravity comprises three distinct contributions \citep{Val01,PC12}; for the electric sail, we have
\begin{equation}\label{EOM}
    \frac{d^2r}{dt^2} = a_{0,E} \left(\frac{d_\star}{r}\right)^{1.25} - g_\star \left(\frac{d_\star}{r}\right)^2 + \frac{\mathcal{L}^2}{r^3},
\end{equation}
where $\mathcal{L}^2 = g_\star d_\star^3$ is the specific angular momentum, while $g_\star$ represents the gravitational acceleration at $d_\star$ and is therefore expressible as
\begin{equation}\label{gstar}
    g_\star = \frac{G M_\star}{d_\star^2} = 5.9 \times 10^{-3}\,\mathrm{m/s^2}\,\left(\frac{M_\star}{M_\odot}\right)^{-2}.
\end{equation}
After multiplying with $\dot{r}$ and integrating (\ref{EOM}) from $r = d_\star$ to $r \rightarrow \infty$, we end up with
\begin{equation}\label{vinfE}
    v_{\infty,E} = \sqrt{v_i^2 + \left(8 a_{0,E} - g_\star\right) d_\star},
\end{equation}
where $v_i$ is the initial velocity imparted to the electric sail at $r = d_\star$. We will set $v_i = 0$ for the time being, although we will later address the possibility of $v_i \neq 0$. In this limit, we find that the terminal velocity of the electric sail becomes
\begin{equation}\label{vtermE}
   v_{\infty,E} \sim 424\,\mathrm{km/s}\,\left(\frac{M_\star}{M_\odot}\right)^{-0.63} \sqrt{1 - 4.9 \times 10^{-3}\left(\frac{M_\star}{M_\odot}\right)^{0.755}}.
\end{equation}
This result is higher by a factor of $3.5$ with respect to Ref. \cite{JA07} for reasons explained in the paragraph immediately after (\ref{AccESail}). 

\begin{figure}
\includegraphics[width=7.5cm]{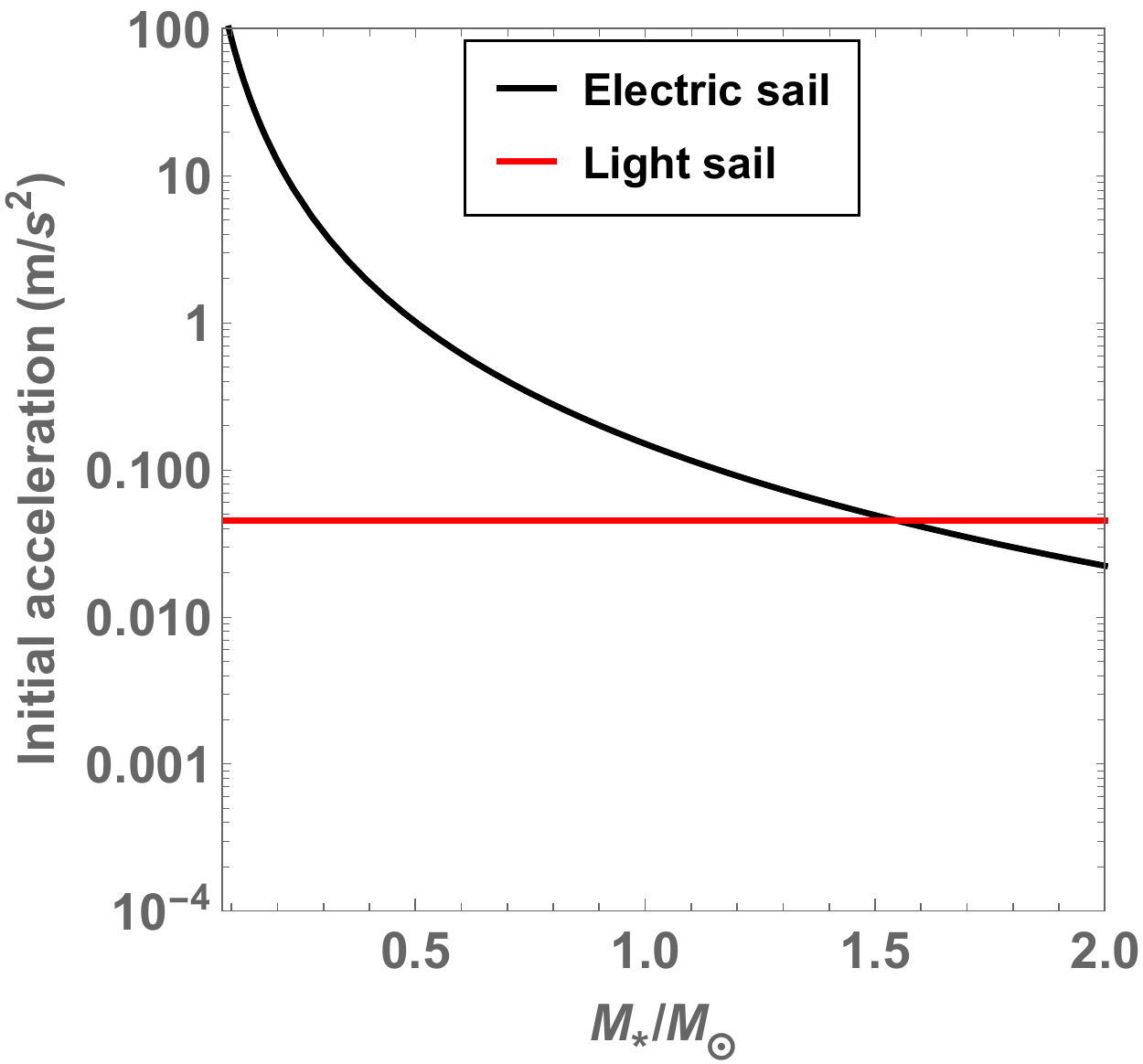} \\
\caption{The initial acceleration experienced by the two propulsion systems as a function of the stellar mass $M_\star$ (in units of $M_\odot$). The parameters for the electric sail were adopted from Ref. \cite{JA07}, while the corresponding choices for the light sail were taken from Ref. \cite{Park18}.}
\label{FigIntAcc}
\end{figure}

For the sake of completeness, we can also calculate the terminal speed achievable if the electric sail is launched from $1$ AU, irrespective of the star's spectral type. While this approach is disadvantageous from the standpoint of attaining higher final velocities, it gives rise to one major benefit - the mass of the spacecraft that must be expended on the electron gun will be reduced significantly for M-dwarfs; this can be verified by comparing $r = 1$ AU with $r = d_\star$ after substituting these choices into (\ref{EGunMass}). In turn, this renders our assumption concerning the payload (being negligible or comparable to sail mass) much more reasonable. In this event, after repeating the same procedure starting with the equation of motion, we obtain
\begin{equation}\label{vtermEAU}
   v_{\infty,E} \sim 424\,\mathrm{km/s}\,\left(\frac{M_\star}{M_\odot}\right)^{-0.44} \sqrt{1 - 4.9 \times 10^{-3}\left(\frac{M_\star}{M_\odot}\right)^{1.88}}.
\end{equation}
Upon comparing (\ref{vtermEAU}) and (\ref{vtermE}), we see that the two expressions are similar, although the latter displays a stronger dependence on $M_\star$ than the former. Let us consider, for instance, a star with $M_\star \approx 0.5 M_\odot$ and assume that an electric sail is launched from $r = d_\star$ and $r = 1$ AU. Using (\ref{vtermE}) and (\ref{vtermEAU}), we estimate the respective terminal speeds to be $\sim 655$ km/s and $\sim 575$ km/s. Thus, not only are final velocities greater than $0.001\,c$ attainable by electric sails in principle near low-mass stars, but also the location (i.e., distance) of the launch site has an ostensibly weak effect on the resulting speeds. 

There is, however, an important caveat that needs to be highlighted at this juncture. Once the electric sail attains speeds that are approximately equal to that of the stellar wind, the velocity $v$ in (\ref{FbyL}) should be replaced by the relative velocity of the sail and the wind. In consequence, when this relative velocity becomes small, the acceleration is diminished greatly and the upper bound on the speed attainable by the electric sail would be effectively set by the speed of the stellar wind \citep{Jan04}. For the parameters adopted in our calculations, this translates to an upper bound of $\sim 500$ km/s for the electric sail; for active stars, this upper bound may increase by a factor of $\gtrsim 2$ due to faster stellar winds and regular coronal mass ejections. Therefore, in conclusion, the terminal velocity of the electric sail is more accurately modelled by $\mathrm{min}\{v_{\infty,E},v_0\}$, where $v_{\infty,E}$ is set by (\ref{vtermE}) or (\ref{vtermEAU}) depending on the context and $v_0$ is the stellar wind velocity introduced previously.

\section{Effectiveness of electric sails as propulsion systems}\label{SecEPS}
Here, we will delve into some of the primary pros and cons of electric sails, especially in comparison to light sails propelled by stellar radiation.

\subsection{Electric sails versus solar sails}\label{SSecLiSa}
Hitherto, we have not undertaken any direct comparison with light sails propelled by stellar radiation. The force exerted by the stellar radiation on the light sail is $F = 2 P_\mathrm{rad} \mathcal{A}$, where $\mathcal{A}$ signifies the area of the light sail, and the factor of $2$ accounts for an ideal light sail with a reflectance of unity \citep{McIn99}. Hence, the acceleration experienced by the light sail ($a_L$) is
\begin{equation}
 a_L = \frac{L_\star}{2 \pi r^2 c \sigma} \sim 4.5 \times 10^{-2}\,\mathrm{m/s^2}\,\left(\frac{M_\star}{M_\odot}\right)^{3}\left(\frac{r}{1\,\mathrm{AU}}\right)^{-2} \left(\frac{\sigma}{\sigma_0}\right)^{-1},
\end{equation}
where $\sigma$ represents the mass per unit area of the light sail. We have introduced the characteristic value of $\sigma_0 \approx 2 \times 10^{-4}$ kg/m$^2$ based on the parameters for the \emph{Breakthrough Starshot} system delineated in Table 1 of Ref. \cite{Park18}. As noted before, we have supposed that the payload mass is negligible (or comparable at most) relative to that of the sail. At the distance $r = d_\star$, we find that the corresponding acceleration (denoted by $a_{0,L}$) simplifies to
\begin{equation}\label{aLint}
    a_{0,L} \sim 4.5 \times 10^{-2}\,\mathrm{m/s^2}\,\left(\frac{\sigma}{\sigma_0}\right)^{-1}.
\end{equation}
Note that $a_{0,L}$ is independent of $M_\star$ because of the scaling $d_\star \propto L_\star^{1/2}$ introduced in (\ref{EqDist}). The initial accelerations imparted to the electric sail and the light sail have been plotted in Fig. \ref{FigIntAcc}. From this plot, it is seen that the light sail attains a higher acceleration than the light sail when $M_\star \gtrsim 1.55 M_\odot$. For smaller masses, our model suggests that electric sails are more suitable for achieving a higher acceleration at the initial location.

\begin{figure}
\includegraphics[width=7.5cm]{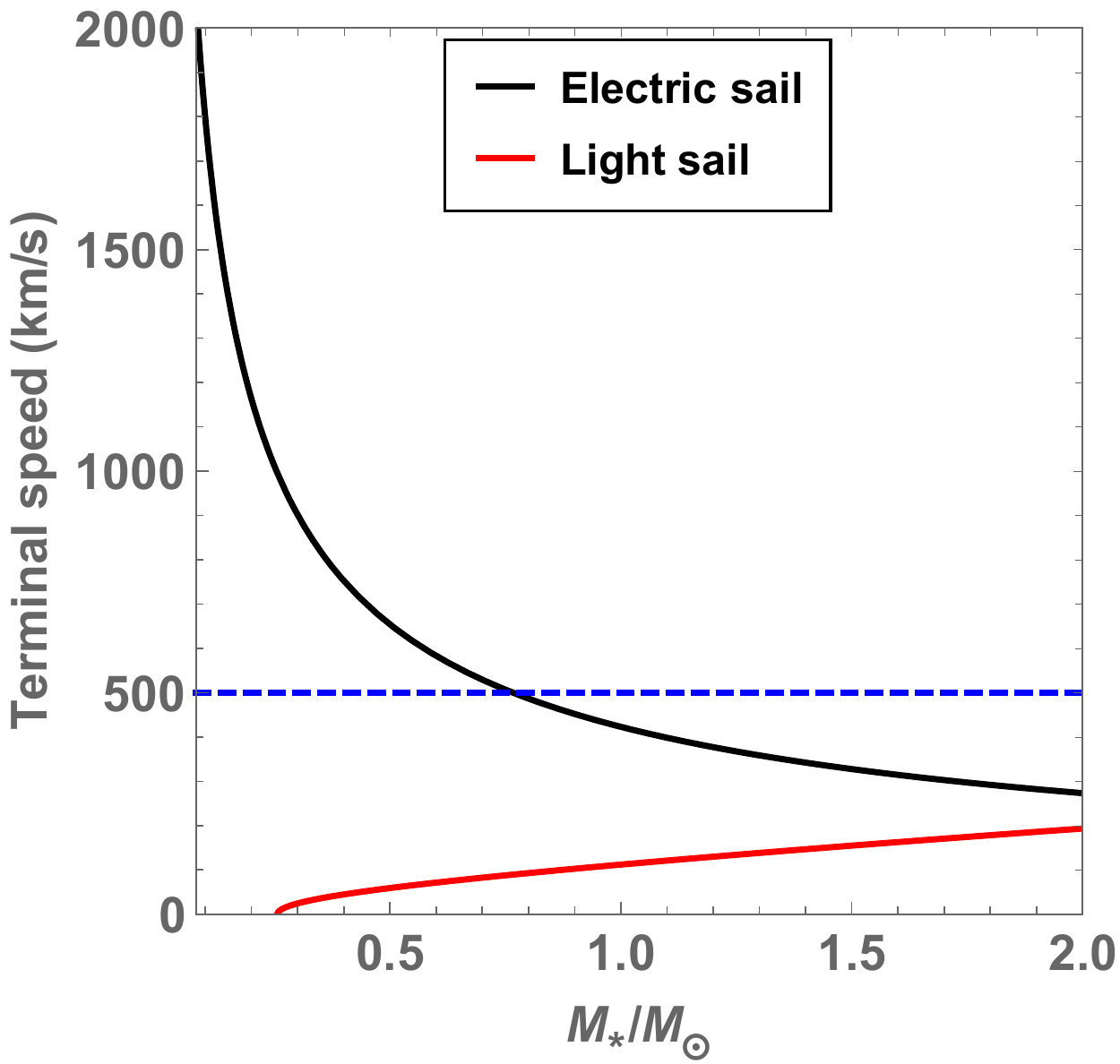} \\
\caption{The terminal speed attained by the two propulsion systems as a function of the stellar mass $M_\star$ (in units of $M_\odot$) assuming they were launched from the distance $d_\star$ given by (\ref{EqDist}). The parameters for the electric sail were adopted from Ref. \cite{JA07}, while the corresponding choices for the light sail were chosen from Ref. \cite{Park18}. The horizontal blue line represents the maximum speed achievable by electric sails, which is effectively set by the speed of the stellar wind.}
\label{FigTermVel}
\end{figure}

We will now calculate the terminal speed attained by the light sail, which is represented by $v_{\infty,L}$. The corresponding equation of motion is
\begin{equation}\label{EOMLS}
    \frac{d^2r}{dt^2} = a_{0,L} \left(\frac{d_\star}{r}\right)^{2} - g_\star \left(\frac{d_\star}{r}\right)^2 + \frac{\mathcal{L}^2}{r^3}.
\end{equation}
By repeating the same procedure undertaken for the electric sail previously, the terminal speed $v_{\infty,L}$ is found to be
\begin{equation}\label{vtLS}
    v_{\infty,L} = \sqrt{v_i^2 + \left(2 a_{0,L} - g_\star\right) d_\star}.
\end{equation}
As before, after supposing that the light sail is launched from rest ($v_i = 0$), we end up with
\begin{equation}\label{vtermL}
   v_{\infty,L} \sim  116\,\mathrm{km/s}\,\left(\frac{M_\star}{M_\odot}\right)^{0.75} \sqrt{\left(\frac{\sigma}{\sigma_0}\right)^{-1} - 0.066 \left(\frac{M_\star}{M_\odot}\right)^{-2}}.
\end{equation}
This expression yields an interesting result. If we specify $\sigma \approx \sigma_0$, it is found that the square root becomes imaginary when $M_\star \lesssim 0.26 M_\odot$. Hence, for M-dwarfs in this mass range, light sails driven by stellar radiation are seemingly not practical in most instances. One can, however, bypass this bottleneck either by launching the light sail from a closer distance than the fiducial choice of $d_\star$,\footnote{If the launch site is very close to the star, the sail incurs the cost of sustained intense heating. Additional orbital maneuvers may also be necessary in order to position the light sail at the initial distance of $r < d_\star$.} imparting a sufficiently high initial velocity $v_i$, or by utilizing materials with $\sigma < \sigma_0$.

We have plotted the terminal speeds achievable by electric sails and light sails as a function of $M_\star$ in Fig. \ref{FigTermVel}. It is apparent from this figure that electric sails permit final speeds of $\sim 500$ km/s to be attained if the stars under question are characterized by $M_\star \lesssim 0.75 M_\odot$; the reason for imposing the cutoff of $\sim 500$ km/s was described in the last paragraph of Sec. \ref{SSecPropES}. Based on Fig. \ref{FigTermVel}, we find that electric sails are rendered more effective than light sails in terms of reaching higher velocities across the entire mass range addressed in this paper. Hence, our analysis suggests that electric sails might represent more efficient alternatives to light sails for F-, G-, K- and M-type stars insofar as achieving high terminal speeds is concerned. 

\begin{figure}
\includegraphics[width=7.5cm]{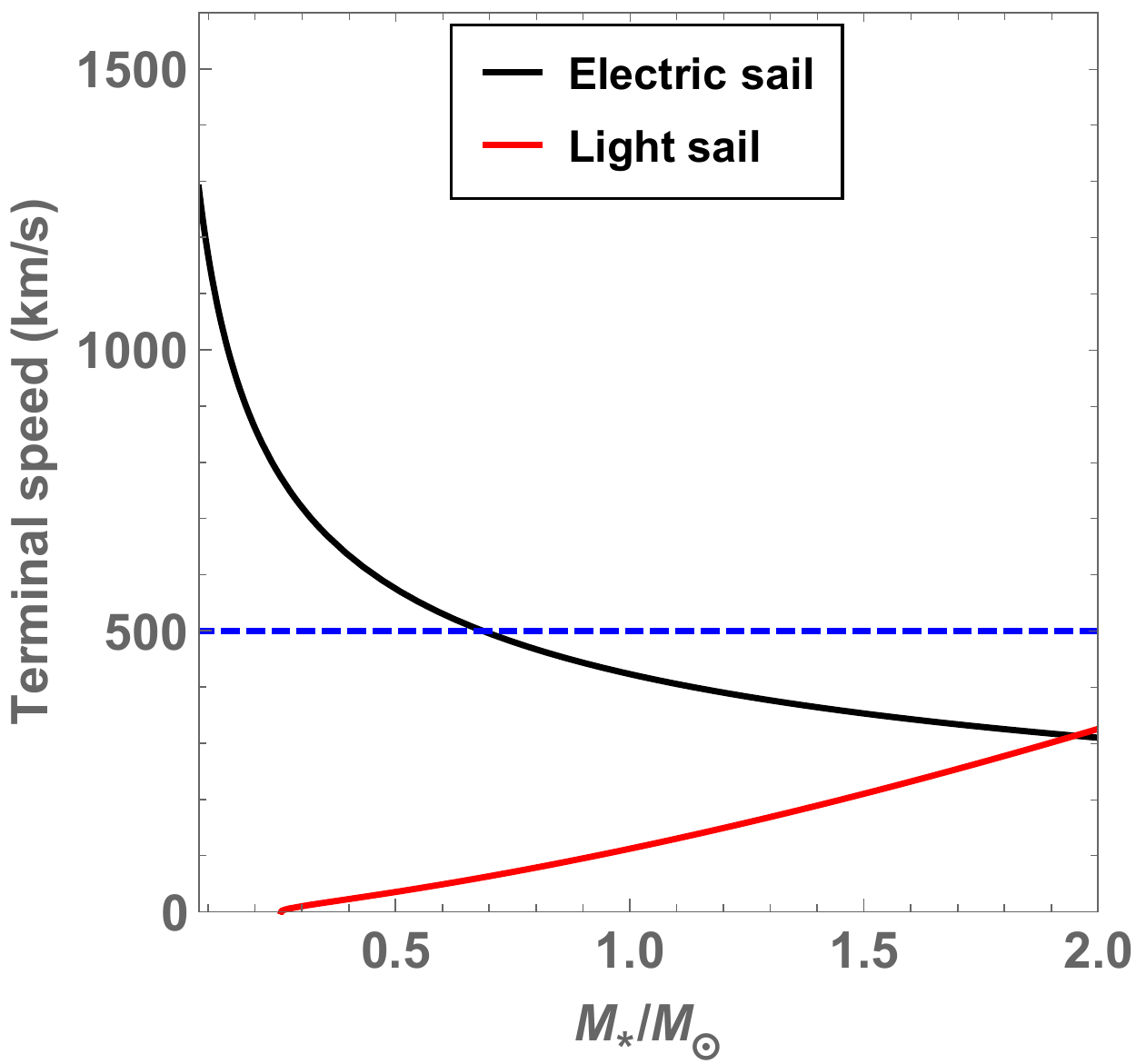} \\
\caption{The terminal speed attained by the two propulsion systems as a function of the stellar mass $M_\star$ (in units of $M_\odot$) assuming they were launched from a fixed distance of $1$ AU. The parameters for the electric sail were adopted from Ref. \cite{JA07}, while the corresponding choices for the light sail were taken from Ref. \cite{Park18}. The horizontal blue line represents the maximum speed achievable by electric sails, which is effectively set by the speed of the stellar wind.}
\label{FigTVelAU}
\end{figure}

For the sake of comparison with (\ref{vtermEAU}), let us suppose that all light sails are launched from a fixed distance of $1$ AU irrespective of spectral type. Upon calculating the terminal velocity following the same procedure as earlier, we arrive at
\begin{equation}\label{vtermLAU}
   v_{\infty,L} \sim  116\,\mathrm{km/s}\,\left(\frac{M_\star}{M_\odot}\right)^{1.5} \sqrt{\left(\frac{\sigma}{\sigma_0}\right)^{-1} - 0.066 \left(\frac{M_\star}{M_\odot}\right)^{-2}}.
\end{equation}
We see that (\ref{vtermLAU}) and (\ref{vtermL}) are akin to each other, except for the fact that the former has a stronger dependence on $M_\star$ than the latter. We have plotted the terminal speeds for electric and light sails, assuming that they are launched from $1$ AU, in Fig. \ref{FigTVelAU}. As per this figure, we determine that final velocities of $\sim 500$ km/s may be achievable by electric sails if the criterion $M_\star \lesssim 0.7 M_\odot$ is fulfilled. We also observe that light sails become faster than electric sails only across the relatively narrow range of $M_\star \gtrsim 1.95 M_\odot$. Thus, even for certain fixed values of the launch distance, electric sails are ostensibly more effective in terms of achieving higher terminal speeds compared to light sails for the majority of F-, G-, K- and M-type stars.

\subsection{Using electric sails and light sails in interstellar travel}
After the electric sail has attained its terminal velocity at a distance far away from the star, it will need to traverse the interstellar medium. There are two distinct possibilities at play here: (i) the interstellar medium can be used for the purposes of deceleration such that the spacecraft is slowed down by the time it reaches the destination, and (ii) the electric sail can be shut down, thus preventing the interstellar medium from decelerating the spacecraft. In the second scenario, the electric sail would travel at a roughly constant speed of $\sim 500$ km/s. It would therefore require $\sim 2 \times 10^3$ yrs to traverse a distance of $\sim 1$ pc. While this timescale is longer than the human lifespan by more than an order of magnitude, this method of propulsion may nevertheless be convenient for transporting larger spacecrafts (e.g., world ships) to neighboring stars across distances of $\sim 1$-$10$ pc.

The first scenario was explored by Ref. \cite{PH16}, who showed that a spacecraft of mass $8250$ kg could be slowed down from $0.05\,c$ to interplanetary speeds in a span of $35$ years by utilizing an electric sail. Due to the complexity of the ensuing model for the deceleration, it is instructive to simplify the problem and obtain a rough estimate of the deceleration that would be experienced by a functional electric sail. As the interstellar medium has multiple phases and is markedly heterogeneous, it is not easy to chose fiducial values for $n$ and $T_e$ \citep{Draine}. Nonetheless, we will adopt $n \sim 10^6$ m$^{-3}$ and $T_e \sim 10^4$ K for the warm interstellar medium. After substituting these values in  (\ref{FbyL}) and using the previous simplifications, we obtain
\begin{equation}
    \frac{dF}{dz} \sim -1.8 \times 10^{-8}\,\mathrm{N/m}\,\left(\frac{v_s}{v_{0,s}}\right)^2,
\end{equation}
where $v_s$ is the instantaneous sail velocity and $v_{0,s}$ signifies the initial sail velocity; the negative sign accounts for the existence of deceleration. As the initial velocity of the sail when it enters the interstellar medium is roughly the same as the terminal velocity when it exits the star's heliosphere, we can set $v_{0,s} = v_{\infty,E}$. If we divide the above equation by (\ref{MeshMass}) and neglect the payload mass (i.e., the electron gun) to maintain consistency with our prior analysis, the ensuing deceleration ($a_\mathrm{ISM}$) is given by
\begin{equation}
    a_\mathrm{ISM} \sim -1.4 \times 10^{-2}\,\mathrm{m/s^2}\,\left(\frac{v_s}{v_{0,s}}\right)^2.
\end{equation}
By integrating this equation, we can determine the time ($\tau$) taken for the spacecraft to slow down until it reaches the requisite speed. After some simplification, we end up with
\begin{equation}
    \tau \sim 57\,\mathrm{yr}\,\left(\frac{v_{\infty,E}}{500\,\mathrm{km/s}}\right)^2 \left(\frac{v_\mathrm{fin}}{10\,\mathrm{km/s}}\right)^{-1},
\end{equation}
where $v_\mathrm{fin}$ is the desired final velocity of the electric sail after it has slowed down; the normalization corresponds to the typical values of $\Delta v$ for interplanetary maneuvers. Hence, at first glimpse, it would seem as though the deployment of an electric sail is efficient at decelerating a spacecraft with negligible payload. There is, however, an important caveat that needs to be recognized here. As the velocity of the sail declines due to momentum exchange with the interstellar medium, the denominator in the right-hand-side of (\ref{FbyL}), which was hitherto neglected, will play an increasingly important role. A more detailed treatment of the deceleration that accounts for a finite payload mass and the denominator from (\ref{FbyL}) can be found in Ref. \cite{PH16}. 

Now, we turn our attention to light sails and explore their feasibility for interstellar travel. During their passage in the interstellar medium, the frictional force is expected to be minimal, i.e., the amount of slowing down can be neglected \citep{HLBL,BL18}. We will now carry out a heuristic calculation for the following scenario. A light sail is launched from star \#1 and enters the interstellar medium. It continues to move at $\sim v_{\infty,1}$, the terminal velocity upon exiting star \#1, until it navigates toward a second star (which is labelled star \#2). While traversing through the interstellar medium and approaching star \#2, we will suppose that the light sail has been retracted or folded to minimize damage. Let us assume that the light sail is unfolded when it reaches a location close to the critical distance $d_\star$ associated with star \#2. By doing so, it will be subject to acceleration as outlined in Sec. \ref{SSecLiSa}. It will thus acquire a final velocity $v_\infty'$ that is estimated using (\ref{vtLS}), thus yielding
\begin{equation}\label{vtwosum}
    v_\infty'^2 = v_{\infty,1}^2 + v_{\infty,2}^2,
\end{equation}
with the terms on the right-hand-side defined to be
\begin{equation}
 v_{\infty,i}^2 = \left(2 a_{i} - g_i\right) d_i
\end{equation}
where $a_i$, $g_i$ and $d_i$ are calculated by substituting the stellar parameters of the $i$-th star into (\ref{aLint}), (\ref{gstar}) and (\ref{EqDist}), respectively; in this equation, note that $i = \{1,2\}$. Alternatively, we can choose a fixed launch distance (such as $1$ AU), in which case the values of $a_i$, $g_i$ and $d_i$ should be recalculated along the lines described in Sec. \ref{SSecLiSa}.

In principle, this mode of operation can be repeated \emph{ad infinitum}. Let us suppose that the total number of ``kicks'' from the radiation pressure of stars encountered by the spacecraft, equipped with an onboard navigation mechanism to accomplish the desired functions, is $N$. The cumulative terminal velocity ($v_c$) after these $N$ kicks becomes
\begin{equation}\label{vcumul}
  v_c^2 =  \sum_{i=1}^N v_{\infty,i}^2.
\end{equation}
This process shares some resemblance with the mechanism of Fermi acceleration that has been widely posited to accelerate cosmic rays at collisionless shocks \citep{BE87,Long11}. Due to the variation of stellar masses - implying that different terminal velocities are associated with each star - analyzing the above expression further is complicated. However, upon inspecting Fig. \ref{FigTermVel} and Fig. \ref{FigTVelAU}, we notice that the terminal velocity is $\sim 100$ km/s for most F-, G- and K-type stars. Moreover, it should be noted that this trio of spectral classes collectively comprise $\sim 20\%$ of all stars in the Milky Way \citep{Krou01}. Thus, we will combine these two facts and assume that all of the target stars chosen by the light sail impart a similar terminal speed to the spacecraft; in quantitative terms, this amounts to selecting $v_{\infty,i} \sim 100$ km/s. With this simplification, the cumulative speed is expressible as
\begin{equation}
    v_c \sim \sqrt{N} v_{\infty,i} \sim 0.11\,c\, \left(\frac{N}{10^5}\right)^{1/2}\left(\frac{v_{\infty,i}}{100\,\mathrm{km/s}}\right)
\end{equation}
Thus, in order to achieve relativistic speeds, the spacecraft would need to reach the vicinity of $\sim 10^5$ stars as per the above equation. At this juncture, we note that the light sail can be decelerated to low speeds over short timescales whenever necessary by coupling it to an electric sail and activating the latter in the interstellar medium as described earlier. 

We can also estimate the total travel time ($t_c$) taken for this final speed to be reached. As per (\ref{vcumul}), the velocity after exiting the first star is $v_{\infty,i} \sim 100$ km/s, the velocity after exiting the second star is $\sqrt{2} v_{\infty,i}$, and so on. Let us employ $\ell$ to denote the average spacing between uniformly distributed target stars. The mid-plane stellar number density in the Milky Way is $\sim 0.1$ pc$^{-3}$, which is derivable from the stellar mass density \citep{Bo17}. Combined with the fact that $\sim 20\%$ of all stars represent viable targets for imparting velocity boosts to the light sail, we adopt a stellar density of $\eta_\star \sim 0.02$ pc$^{-3}$. \footnote{\url{http://www.pas.rochester.edu/~emamajek/memo_star_dens.html}} Hence, $\ell$ is duly determined by demanding that $4\pi \eta_\star \ell^3/3 \sim 1$, which yields $\ell \sim 2.3$ pc after simplification. Thus, in accordance with the above assumptions, the total travel time will be
\begin{equation}
    t_c \approx \frac{\ell}{v_{\infty,i}} + \frac{\ell}{\sqrt{2} v_{\infty,i}} + \dots \frac{\ell}{\sqrt{N-1} v_{\infty,i}} \approx \frac{\ell}{v_{\infty,i}} H_{N-1}^{(1/2)},
\end{equation}
where $H_n^{(m)}$ is the $n$-th generalized harmonic number of order $m$, whose properties are well understood \citep{SC01}. As we are typically interested in large $N$, the above formula is simplified by making use of the asymptotic expansion $H_n^{(1/2)} \sim 2\sqrt{n}$ for large $n$, which emerges upon applying the Euler–Maclaurin formula \citep{AS65}. Thus, when the limit $N \gg 1$ is calculated for the above equation, we obtain
\begin{equation}
   t_c \sim 14\,\mathrm{Myr}\,\left(\frac{N}{10^5}\right)^{1/2}\left(\frac{v_{\infty,i}}{100\,\mathrm{km/s}}\right)^{-1}. 
\end{equation}
Although $\sim 10$ Myr represents a long timescale from the perspective of the lifetime associated with a particular species, it is nevertheless small in comparison to the age of the Milky Way. Furthermore, the average age of terrestrial planets in the Milky Way is a few Gyr older than the Earth \citep{Line01}. Thus, in theory, any putative technological species - irrespective of whether they are biological or post-biological in nature \citep{Dick08} - that evolved before \emph{Homo sapiens} would have had plenty of time to undertake the requisite number of encounters to achieve relativistic speeds via light sails. Once this objective has been accomplished, interstellar exploration of the remaining regions of the Milky Way could be undertaken at $v_c$. Alternatively, due to the fact that $v_c$ is relativistic, it may be possible to undertake intergalactic travel and reach other galaxies in the Local Group. Assuming that humans navigate the existential risks of the Anthropocene and subsequent epochs successfully, it is conceivable that the descendants of \emph{Homo sapiens} might also opt for this mode of interstellar exploration.

\section{Conclusions}\label{SecConc}
We investigated the relative effectiveness of light sails powered by stellar radiation versus electric sails driven by stellar winds. We analyzed how the basic features of these two propulsion systems depend on stellar parameters (mass and rotation rate). To simplify our treatment, we chose to vary only the stellar mass and focused on F-, G-, K- and M-type stars, the last of which are the most abundant in the Milky Way. 

We began by showing that the stellar wind pressure may dominate over the radiation pressure when the stellar mass obeys $M_\star \lesssim 0.2 M_\odot$. Subsequently, after calculating the initial acceleration and terminal speeds attained by the two propulsion systems, we showed that electric sails are likely to be more effective than light sails for the majority of F-, G-, K- and M-type stars. We found that the final velocities attained by electric sails (with minimal payload masses) do not exhibit a strong dependence (in certain cases) on the initial location from where they are launched. We also demonstrated that speeds of $\sim 500$ km/s are potentially achievable via electric sails near K- and M-dwarfs. With regards to late-type M-dwarfs, they incur the potential disadvantage of not having sufficient radiation pressure to counterbalance the gravitational acceleration exerted by the host star, thus rendering light sails propelled by stellar radiation impractical. 

Next, we studied the issue of deceleration engendered by the interstellar medium and showed that it can contribute significantly to the slowing down of electric sails. However, in the event that the electron gun is turned off in the interstellar medium, electric sails represent a promising avenue for undertaking interstellar exploration of stellar systems in the neighborhood of the origination point (i.e., at distances of $\sim 1$-$10$ pc). Subsequently, we analyzed the prospects for interstellar travel by means of light sails driven by stellar radiation. We proposed a strategy by which such spacecraft can achieve progressively higher speeds via a series of repeated encounters with F-, G- and K-type stars.\footnote{The increments arising from stellar radiation pressure may be supplemented by resorting to conventional orbital maneuvers such as gravitational slingshots and Oberth maneuvers.} We showed that sampling $\sim 10^5$ stars could enable light sails to achieve relativistic speeds of $\sim 0.1\,c$ and that this mechanism would require $\sim 10$ Myr. While this constitutes a long timescale by human standards, it is not especially long in comparison to many astronomical and geological timescales. The ensuing relativistic light sails would be well-suited for engaging in interstellar exploration and even voyaging to neighboring galaxies.\footnote{Relativistic spacecraft are also useful from the standpoint of executing directed panspermia efficiently \citep{Lin16,LM16}, i.e., deliberately seeding other worlds with life \citep{CO73}.}

Aside from the advantages outlined until now, there are several other benefits accruing from the deployment of electric sails. First, as the electric sail comprises a wire mesh, it has a much lower cross-sectional area with respect to a light sail of the same overall dimensions. Hence, it is relatively less susceptible to the deleterious effects of coronal mass ejections and stellar proton events near the star as well as dust grains and cosmic rays in the interstellar medium. Second, due to the low cross-sectional area, modest dimensions of putative electric sails and the natural variability of stellar winds, it seems very unlikely that their presence will be readily detectable by the same methods proposed for exoplanets, such as radio emission via the electron-cyclotron maser instability \citep{Zar07}. On the whole, there are compelling grounds for supposing that technologically sophisticated species might opt to use this propulsion mechanism, especially if they wish to minimize their likelihood of being detected.\footnote{This is a highly non-trivial question for humanity because there has been much debate as to whether deliberately broadcasting our presence would be detrimental or beneficial to us \citep{Mus12,HBS,Gertz}.}

While our analysis supports the notion that electric sails are more advantageous vis-\`a-vis light sails for interplanetary and short-term interstellar travel, it is worth emphasizing that we are only dealing with light sails powered by stellar radiation. In contrast, as the ongoing example of \emph{Breakthrough Starshot} illustrates, one can instead utilize a laser array to accelerate light sails to relativistic speeds. Thus, for species with the objective of attaining relativistic spaceflight over a short time span, thereby obviating the necessity of embarking on a long-term sequence of velocity boosts, laser-driven light sails may prove to be an optimal solution. However, the downsides to using lasers to propel light sails are: (i) it is energetically very demanding, and (ii) it runs the risk of signalling the presence of the technological species in question \citep{GuLo,LL17}.

In summary, we studied the feasibility of using electric sails and light sails driven by stellar winds and radiation, respectively. Our analysis indicated that electric sails represent a more effective and promising alternative to light sails, especially insofar as interplanetary travel is concerned. However, as our analysis neglected some notable stellar parameters (e.g., age-dependent rotation rate) and did not take a number of real-world engineering specifications and issues into account, further work is necessary in order to investigate and demonstrate the feasibility of electric sails as propulsion systems in the vicinity of low-mass stars.

\section*{Acknowledgments}
We thank the reviewers for their helpful feedback, which was helpful for improving the paper. This work was supported in part by the Breakthrough Prize Foundation, Harvard University's Faculty of Arts and Sciences, and the Institute for Theory and Computation (ITC) at Harvard University.


\end{document}